\documentclass[12pt]{article}
\usepackage{epsfig}

\newcommand{\be}{\begin{equation}}
\newcommand{\ee}{\end{equation}}
\newcommand{\bea}{\begin{eqnarray}}
\newcommand{\eea}{\end{eqnarray}}
\newcommand{\nn}{\nonumber}
\newcommand{\al}{\alpha}

\newcommand{\G}{\Gamma}

\newcommand{\MSsch}{{\overline{\rm MS}}}
\newcommand{\ice}[1]{\relax}

\newcommand{\ba}{\begin{array}} 
\newcommand{\ea}{\end{array}}

\newcommand{\Li}{{\rm Li}}

\newcommand{\pfrac}[2]{\left(\frac{#1}{#2}\right)}
\begin{document}
\thispagestyle{empty}
\begin{flushright}
MZ-TH/00-40\\
September 2000\\
\end{flushright}
\vspace{0.5cm}
\begin{center}
{\Large\bf High order PT calculations for heavy quarks\\
near threshold}\\[2truecm]
{\large \bf A.A.Pivovarov}\\[0.5truemm]
Institut f\"ur Physik, Johannes-Gutenberg-Universit\"at,\\
Staudinger Weg 7, D-55099 Mainz, Germany\\
and \\
Institute for Nuclear Research of the\\
Russian Academy of Sciences, Moscow 117312, Russia
\end{center}

\vspace{2truecm}
\begin{abstract}
Results of analytical calculations for 
heavy quark systems in higher orders of perturbation theory are
overviewed. I discuss baryons with one finite mass quark
in the next-to-leading order and heavy quark pair production
near the threshold within NRQCD.
\end{abstract}

\vskip 3.5cm
\begin{center}
{\it Talk given at  Research Workshop
``Calculations for Modern and Future Colliders'' \\

JINR, Dubna, Russia, July 9-23, 2000}
\end{center}

\newpage
\section{Introduction.}
Recent years have witnessed some progress in 
theoretical description of heavy
quark systems near the threshold
which will be experimentally studied at future colliders \cite{exp}. 
It is mostly a technical progress
within general approach of effective theories
\cite{weinberggasserleutw}.
The computational difficulties of high order PT calculations
have led to the necessity of introducing some approximations 
realized in the framework of the effective theories. 
Processes with a low momentum transfer for the particles 
containing one heavy quark are well approximated 
by HQET \cite{shifmanvoloshinpolitzWEinGeorNeub}.
This approach was quite successful in describing transitions
between heavy particles containing $b$- and $c$-quarks
\cite{isgurwise}.
For the systems of two heavy particles near the threshold 
the effective theory is NRQCD \cite{CasLep}.
Recent result obtained in its framework is
the completion of NNLO analysis of heavy quark 
production e.g. \cite{revHoa}.
In my talk I briefly present the results of computing 
perturbative corrections
to physical quantities relevant for description of baryons with
one heavy quark
and of the heavy quark pair production near the threshold.

\section{Baryons with one heavy quark.}
Properties of the spectrum of heavy baryons can be obtained from the
analysis of the correlator of two baryonic currents.
Calculations with massive quarks are basically 
done in the leading order only. The NLO corrections in the 
massless case have been known since long ago \cite{pivbar}.
The generic baryonic current is a local three-quark operator
$j=\epsilon^{abc}(u_b^T C d_c)\G \Psi_a $
where $\Psi$ is a heavy quark with the mass $m$, $u,d$ are massless quarks,
$C$ is the charge conjugation matrix, $\G$ is a Dirac matrix.
$\epsilon^{abc}$ is the totally antisymmetric 
tensor, $a,b,c$ are indices for the $SU(3)$ color group.
Both functions $\Pi_{q,m}(q^2)$ of the correlator 
of two baryonic currents (for simplicity we take $\G=1$)
\be
\label{def00}
i\int \langle T j(x) \bar j (0) \rangle e^{i q x}dx
=\gamma_\nu q^\nu \Pi_q(q^2)+ m \Pi_m(q^2)
\ee
are known up to NLO of PT (three-loop diagrams). 
I shall present results only for 
the function $\Pi_m(q^2)$ \cite{grobar}.
For the spectral density $\rho(s)$ in the dispersion relation 
\be 
\Pi_m(q^2)=\frac{1}{128\pi^4}\int_{m^2}^\infty \frac{\rho(s)ds}{s-q^2}
\ee 
we write
\be
\rho(s)=s^2\left(\rho_0(s)
(1+\frac{\al_s}{\pi}\ln\frac{\mu^2}{m^2})
+\frac{\al_s}{\pi}\rho_1(s)\right)\, .
\ee
Here $\mu$ is the scale parameter, $\al_s=\al_s(\mu)$,
$m$ is a pole mass of heavy quark \cite{polemass}.
The leading order term reads
\be
\label{lead0}
\rho_0(s)=1+9z-9z^2-z^3+6z(1+z)\ln z 
\ee
with $z=m^2/s$. In the $\MSsch$-scheme the NLO term is given by 
\begin{eqnarray}\label{corr1}
\lefteqn{\rho_1(s)\ =\ 9+\frac{665}9z-\frac{665}9z^2-9z^3}\\&&
  -\left(\frac{58}9+42z-42z^2-\frac{58}9z^3\right)\ln(1-z)
  +\left(2+\frac{154}3z-\frac{22}3z^2-\frac{58}9z^3\right)\ln z\nonumber\\&&
  +4\left(\frac13+3z-3z^2-\frac13z^3\right)\ln(1-z)\ln z
  +12z\left(2+3z+\frac19z^2\right)\left(\frac12\ln^2z-\zeta(2)\right)
  \nonumber\\&&
  +4\left(\frac23+12z+3z^2-\frac13z^3\right)\Li_2(z)
  +24z(1+z)\left(\Li_3(z)-\zeta(3)-\frac13\Li_2(z)\ln z\right)\nonumber
\end{eqnarray}
where $\Li_n(z)$ are polylogarithms and $\zeta(n)$ is Riemann's zeta
function. The diagrams have generalized sunset
topology which has been recently studied in some detail
\cite{sun0}. Such a topology of diagrams is 
rather frequent in phenomenological applications \cite{sunappl}.
Since the anomalous dimension of the baryonic current
is known to two-loop order \cite{pivan}
the result eq.~(\ref{corr1}) completes the NLO analysis.

Two limiting cases of interest
are the near-threshold and high-energy asymptotics.
With eq.~(\ref{corr1}) both limits can be taken 
explicitly. Effective theories can be viewed as 
special devices for calculations in such limiting cases.

In the high energy (small mass) limit $z\rightarrow 0$
the correction reads
\be
\label{corr12}
\rho_1(s)= 9 + 83\,z - 4\,{{\pi }^2}\,z 
+ 2\,\ln (z) + 50\,z\,\ln(z) 
+ 12\,z\,{{\ln (z)}^2} - 24\,z\,\zeta(3)+O(z^2)\, .
\ee
This leads to the small mass expansion 
of the spectral density in the form
\be
\label{m00}
m\rho(s)=m_{\MSsch}(\mu) \rho_{massless}(s)=
m_{\MSsch}(\mu)s^2 \left(1+\frac{\al_s}{\pi}\left(
2\ln\frac{\mu^2}{s}+\frac{31}{3}\right)
\right)
\ee
where $\rho_{massless}(s)$
is the result of calculating the correlator in the effective theory 
of massless quarks.
The relation between the pole 
mass $m$ and the $\MSsch$ mass $m_{\MSsch}(\mu)$ reads
\be
m=m_{\MSsch}(\mu)\left(1+\frac{\al_s}{\pi}\left(
\ln\frac{\mu^2}{m^2}+\frac{4}{3}\right)
\right) \, .
\ee
The massless effective theory cannot reproduce 
the mass singularities -- terms like $z\,\ln(z)$
in eq.~(\ref{corr12}). Within the massless effective theory
such contributions can be parametrized 
with condensates of local operators. The first $m^2$
correction in eq.~(\ref{corr12})
can be found if the perturbative value of the heavy quark condensate
$\langle \bar \Psi  \Psi\rangle$ 
taken from the full theory is added.
The vacuum expectation value $\langle \bar \Psi  \Psi\rangle$
of the composite operator 
$(\bar \Psi  \Psi)$ should be understood within a mass independent 
renormalization scheme such as the $\MSsch$-scheme.
This value (perturbative, $\langle \bar \Psi  \Psi\rangle\sim m^3
\ln(\mu^2/m^2)$) cannot be computed 
within the effective theory of massless quarks.
It represents the proper matching between 
the perturbative expressions for the correlators
of the full (massive) and effective (massless) theories.
This matching procedure allows one to restore 
higher order terms of the mass expansion 
in the full theory from the effective massless theory
with the mass term treated as a perturbation.

In the near-threshold limit $E\rightarrow 0$
with $s=(m+E)^2$
one explicitly obtains
\be
\label{hqet0}
\rho_{th}(m,E)=\frac{16 E^5}{5m}\left(1+\frac{\al_s}{\pi}\ln\frac{\mu^2}{m^2}
+\frac{\al_s}{\pi}\left(\frac{54}{5}+\frac{4\pi^2}{9}
+4\ln\frac{m}{2E}\right)\right)+O(E^6)\, .
\ee
In this region the appropriate device to compute 
the limit of the correlator is HQET. Writing
\be
\label{hqet}
m \rho_{th}(m,E)=C(m/\mu,\al_s)^2\rho_{HQET}(E,\mu)
\ee
we obtain the known result
for $\rho_{HQET}(E,\mu)$ \cite{groote} 
\be
\label{explrhoHQET}
\rho_{HQET}(E,\mu)
=\frac{16 E^5}{5}\left(1+\frac{\al_s}{\pi}
\left(\frac{182}{15}+\frac{4\pi^2}{9}+4\ln\frac{\mu}{2E}\right)\right)
+O(E^6)
\ee
with matching coefficient $C(m/\mu,\al_s)$
\cite{barmatch}
\be
\label{matchcoef}
C(m/\mu,\al_s)=1+\frac{\al_s}{\pi}
\left(\frac{1}{2}\ln\frac{m^2}{\mu^2}
-\frac{2}{3}\right)\, .
\ee
The matching procedure allows one to restore the near-threshold limit
of the full QCD correlator (\ref{hqet0})
starting from the result obtained in a simpler HQET
valid near the threshold only.
Higher order terms of $E/m$ expansion
can be easily obtained from the explicit result eq.~(\ref{corr1}).
The NLO correction in low energy expansion reads
\begin{equation}\label{thre1}
\Delta \rho_{th}(m,E)=-\frac{88E^6}{5m^2}\left\{1+\frac{\alpha_s}\pi
  \left(\ln\pfrac{\mu^2}{m^2}+\frac{376}{33}+\frac{4\pi^2}9
  +\frac{140}{33}\ln\pfrac{m}{2E}\right)\right\}\, .
\end{equation}
To obtain this result starting from HQET
is a more difficult task requiring the analysis of contributions
of higher dimension operators. 

Note that the simple interpolation between the massless
and HQET limits gives a rather good approximation of the  
ratio $\rho_1(s)/\rho_0(s)$ for all $m^2<s$. 
For the moments of the spectral density
\be
\label{mombar0}
{\cal M}_n=\int_{m^2}^\infty \frac{\rho(s)ds}{s^n}
=m^{6-2n}M_n , 
\quad M_n=M_n^{(0)}
\left(1+\frac{\al_s}{\pi}(\ln\frac{\mu^2}{m^2}+\delta_n)\right)
\ee
one finds
\be
M_n^{(0)}=\frac{12}{n(n-1)^2(n-2)^2(n-3)} ,
\quad 
\delta_n=A_n+\frac{4}{3}\zeta(2)=A_n+\frac{2\pi^2}9\, .
\ee
The coefficients $A_n$ are rational numbers,
$A_4=3$, $A_5=13/2$, $A_6=17/2$, $A_7=535/54$ \cite{grobar}. 
The expression for
$\delta_n$ with arbitrary $n$ is long. 

\section{Two heavy quarks near threshold.}
With two heavy quarks involved in a physical 
process it is becoming more difficult to perform PT calculations.
Still many results in finite order PT 
for vector current correlators are known
both analytically and numerically \cite{chetkwiaschw3}.
For applications near the threshold 
where $v=\sqrt{1-4m_t^2/s}\ll 1$ ($m_t$ is a generic notation for 
the heavy quark mass) the ordinary perturbation theory
(with free quarks as the lowest order approximation) breaks down.
The ratio $\al_s/v$ is not small
and all terms of the order $(\al_s/v)^n$ should be summed.
The motion of the heavy quark-antiquark pair
near the production threshold is nonrelativistic to 
high accuracy that justifies the use 
of nonrelativistic quantum mechanics
for describing such a system \cite{CasLep}. 
Being much simpler than the comprehensive relativistic treatment 
of the bound state problem
with Bethe-Salpeter amplitude \cite{BetheSal},
this approach allows one to take into account 
Coulomb interaction exactly \cite{VL,ttee}.
However, for this approximation to work the full QCD 
should first be mapped onto NRQCD which is done in PT.
For the analysis of heavy quark production by the vector current
the basic quantity is the correlator
\be
\label{corem}
\Pi(E)=i \int \langle T j_{em}(x)j_{em}(0)\rangle e^{iqx} dx, 
\quad q^2=(2m_t+E)^2\, .
\ee
Near the threshold (for small energy $E$) NRQCD is used.
Up to the NNLO accuracy in NRQCD one has (cf. eq.~(\ref{hqet}))
\be
\Pi(E)=\frac{2\pi}{m_t^2}C_h(\al_s)C_{\cal O}(E/m_t) G(E;0,0)\, .
\label{Rv}
\ee 
Here $C_h(\al_s)$ is the high energy coefficient
(analog of eq.~(\ref{matchcoef})) which has been
known in the NLO since long ago \cite{KarHarBr}.
$G(E;0,0)$ is the nonrelativistic Green's function
(an object of the effective theory as $\rho_{HQET}(E,\mu)$
in eq.~(\ref{explrhoHQET})), $E=\sqrt{s}-2m_t$.
Up to NNLO of NRQCD the contribution
of higher dimension operators can be written as a total factor
$C_{\cal O}(E/m_t)=1-4E/3m_t$ \cite{hoa:match}.
High energy coefficient $C_h(\al_s)$ is given by the expression
\be
\label{hic}
C_h(\al)=1-4\frac{\al_s}{\pi}+C_F\left(\frac{\al_s}{\pi}\right)^2
\left(-\frac{35}{9}\pi^2 \ln\frac{\mu_f}{m_t}+c_2\right)\, .
\ee
The most difficult part of eq.~(\ref{hic}) to obtain
is the coefficient $c_2$ \cite{highcoef}
\[
c_2=\frac{103}{36}+\frac{1003}{216}\pi^2 - \frac{56}{9} \pi^2 \ln 2
-\frac{125}{6}\zeta(3)+\frac{11}{18}n_f ,
\]
$n_f$ is the number of light quarks.
An explicit dependence of high and low energy quantities
on the factorization scale $\mu_f$ 
is a general feature of effective theories which are  
valid only for a given region of energy.
A physical quantity, which is given 
by a proper combination of results obtained 
in different energy regions within respective 
approximations, is factorization scale independent
e.g. \cite{pivKK}. The basic dynamical 
quantity in the analysis of the near-threshold effects is
the nonrelativistic Green's function $G(E)=(H-E)^{-1}$ where 
\be
\label{hamNR}
H=\frac{p^2}{m_t}+V(r)
\ee
is the nonrelativistic Hamiltonian.
A part of Hamiltonian (\ref{hamNR}) which is difficult 
to find is the heavy quark static potential $V_{pot}(r)$ entering 
into the potential $V(r)$. 
The static potential $V_{pot}(r)$ is computed in 
perturbation theory 
\bea
\label{stapot}
V_{pot}(r)=-C_F\frac{\al_s}{r}\left(1
+\frac{\al_s}{4\pi}\left(C_1^1\ln(\mu r)+C_0^1\right)
+\left(\frac{\al_s}{4\pi}\right)^2
\left(C_2^2\ln^2 (\mu r) +C_1^2\ln (\mu r)+C_0^2\right)\right) .
\eea
Nontrivial coefficients of this expansion
are $C_0^1$ and $C_0^2$.
The numerical value for $C_0^1$ was obtain long ago \cite{Fish},
while $C_0^2$ has been recently computed \cite{PeterSchr}.
In order to exactly take into account the Coulomb effects near 
the threshold the Hamiltonian (\ref{hamNR})
is represented in the form \cite{AshBerLandau}
\be
H=H_C+\Delta H, \quad H_C=\frac{p^2}{m_t}-C_F\frac{\al_s}{r}
\ee
with 
\be
\label{dH}
\Delta H = \Delta V_{pot}-\frac{H_C^2}{4m_t}
-\frac{3C_F\al_s}{4m_t}\left[H_C,\frac{1}{r}\right]_{+}
-\frac{4\pi\al_s}{m_t^2}
\left(\frac{C_F}{3}+\frac{C_A}{2}\right)\delta(\vec{r}) \, .
\ee
For QCD one has $C_A=3$, $C_F=4/3$. 
Constructing the Green's function is straightforward and can be done
analytically within perturbation theory 
in $\Delta H$ near the Coulomb Green's 
function $G_C(E)$ \cite{KPP,PPres,BSS1}
or numerically by solving the Schr\"odinger equation
with Hamiltonian (\ref{hamNR}) (for complex values of $E$ 
that can be used to describe the production of particles 
with nonzero width) \cite{Hoang,Mel,Nag,direg,Yak,HoaTeu}.
I'll discuss the analytical solution only.
Some terms in eq.~(\ref{dH}) can be handled efficiently.
The shift of the parameters 
of the Coulomb Green's function $E\to E+E^2/4m_t$
and $\al_s\to \al_s(1+3E/2m_t)$
accounts for the relativistic $H_C^2$ and anticommutator corrections 
in eq.~(\ref{dH}). 
In this respect the modified Coulomb
approximation can be used as the leading order approximation. 
The Hamiltonian can be written in the form 
\be
H=H_{LO}+\Delta V_{pot}+ \al_s V_0\delta(\vec{r}), 
\quad V_0=-\frac{4\pi}{m_t^2}\left(\frac{C_F}{3}+\frac{C_A}{2}\right)
=-\frac{2\pi}{m_t^2}\frac{35}{9}
\ee
where $H_{LO}$ is the (modified) Coulomb approximation,
$\Delta V_{pot}$ is 
the perturbation theory correction to $V_{pot}(r)$.
The $\delta(\vec{r})$-part is a separable potential and can be taken 
into account exactly \cite{nonrelPiv}. The solution reads
\be
G(E;0,0)=\frac{G_{ir}(E;0,0)}{1+\al_s V_0 G_{ir}(E;0,0)}\, ,
\quad
G_{ir}(E)=(H_{LO}+\Delta V_{PT}-E)^{-1} \, .
\ee
Therefore one is left only with the necessity to construct corrections due
to $\Delta V_{pot}$-part which modify $G_{ir}(E)$.
An efficient calculational framework for perturbation theory 
in $\Delta V_{pot}$ near the Coulomb Green's function $G_C(E)$
has been developed \cite{KPP}. The key point is the use of 
a partial wave decomposition for the Green's function
\be
G(E;{\bf x},{\bf y})=\sum^\infty_{l=0}(2l+1)
(xy)^lP_l({\bf xy}/xy)G_l(E;x,y)
\label{lexp}
\ee
where $P_l(z)$ is a Legendre polynomial.
The Coulomb partial waves $G_l^C({\bf x},{\bf y},k)$ are known
\be
G^C_l(E;x,y)={m_t k\over 2\pi}(2k)^{2l}e^{-k(x+y)}
\sum_{m=0}^\infty {L_m^{2l+1}(2kx) L_m^{2l+1}(2ky)m!\over
(m+l+1-\nu)(m+2l+1)!}
\label{Gcl}
\ee
where $k^2=-m_t E$, $\nu=\lambda / k$, $\lambda =\alpha_s C_F m_t/2$,
$L^\al_m(z)$ is a Laguerre polynomial.
The result of the evaluation of the NLO correction in S-wave is
\[
\Delta_1G_0(E)
={\al_s\beta_0\over 2\pi}{\lambda m_t\over 2\pi}\left(
\sum_{m=0}^\infty F(m)^2(m+1)
\left(L_1(k)+\Psi_1(m+2)\right)-2\sum_{m=1}^\infty\sum_{n=0}^{m-1}
F(m)
\right.
\]
\[
\left. \times F(n){n+1\over m-n} +2\sum_{m=0}^\infty F(m)
\left(L_1(k) - 2\gamma_E-\Psi_1(m+1)\right)
-\gamma_E L_1(k)+{1\over 2}L_1(k)^2
\right)
\]
where $\Psi_n(x)=d^n \ln \G(x)/dx^n$, $\beta_0$ is the first
coefficient of the $\beta$-function.
Here 
\[
L_1(k)=\ln\left({\mu_se^{C_0^1/C_1^1}\over 2k }\right)\, ,
\quad F(m)={\nu\over (m+1)\left(m+1-\nu\right)}\, .
\]
The NLO correction to the $l=1$ partial wave
was found in \cite{PP2}
\[
\Delta_1 G_1(E)
={\al_s\beta_0\over 2\pi}{\lambda m_tk^2\over 18\pi}\left(
\sum_{m=0}^\infty \tilde F(m)^2(m+1)(m+2)(m+3)
\left(L_1(k)+\Psi_1(m+4)\right)
\right.
\]
\[
-2\sum_{m=1}^\infty\sum_{n=0}^{m-1}
\tilde F(m)\tilde F(n){(n+1)(n+2)(n+3)\over m-n} +2\sum_{m=0}^\infty
\tilde F(m)\bigg(2\tilde J_0(m)+(m+1)(m+2)L_1(k)
\]
\[
\left.
+(1+\nu)(\tilde J_1(m)+(m+1)L_1(k))
+{\nu(\nu+1)\over 2}(\tilde J_2(m)+
2L_1(k))\bigg)
+\tilde I(k) \right)
\]
where
\[
\tilde F(m)={\nu(\nu^2-1)\over (m+2-\nu)(m+1)(m+2)(m+3)}\, ,
\]
\[
\tilde J_0(m)=-2\Psi_1(m+1)-4\gamma_E +3,
\]
\[
\tilde J_1(m)=(m+1)(-\Psi_1(m+2)-2\gamma_E +2) ,
\]
\[
\tilde J_2(m)={(m+1)(m+2)\over 2}\left(-\Psi_1(m+3)
-2\gamma_E +{3\over2}\right) ,
\]
\[
\tilde I(k)=-{(\gamma_E-1)^2\over 2}-{\pi^2\over 12}
-(4-3\gamma_E)\nu +{1-9\gamma_E+6\gamma_E^2+\pi^2\over 4}\nu^2+
{1-3\gamma_E\over 2}\nu^3+{1-\gamma_E\over 4}\nu^4
\]
\[
+\left(\gamma_E-1-3\nu+{9-12\gamma_E\over 4}\nu^2
+{3\over 2}\nu^3+{1\over 4}\nu^4\right)L_1(k)
+\left(-{1\over 2}+{3\over 2}\nu^2\right)L_1(k)^2 \, .
\]
The Green's function at the origin can be written in the form 
with single poles only
\be
G(E;0,0)=\sum_{m=0}^\infty{|\psi_{0m}(0)|^2\over E_{0m}-E}+
{1\over \pi}\int_0^\infty{|\psi_{0E'}(0)|^2\over E'-E}
dE'
\label{endenom}
\ee
where $\psi_{0m,E}(0)$ is the wave function at the origin.
Up to NNLO one writes
\bea
E_{0m}&=&E^C_{0m}\left(1+\Delta_1E_{0m}+\Delta_2E_{0m}\right) ,\nn \\
|\psi_{0m}(0)|^2&=&|\psi^C_{0m}(0)|^2
\left(1+\Delta_1\psi^2_{0m}+\Delta_2\psi^2_{0m}\right)
\eea
where
\[
E^C_{0m}=-{\lambda^2\over m_t(m+1)^2}, \qquad
|\psi^C_{0m}(0)|^2={\lambda^3\over \pi(m+1)^3}\, .
\]
The analytical expression for the NLO corrections
to the bound state parameters has the form
\[
\Delta_1E_{0m}={\alpha_s\beta_0\over \pi}\left(\bar L_1(m)+\Psi_1(m+2)\right),
\]
\[
\Delta_1\psi^2_{0m}={\alpha_s\beta_0\over 2\pi}
\left(3\bar L_1(m)+\Psi_1(m+2)-2(m+1)\Psi_2(m+1)
-1-2\gamma_E+{2\over m+1}
\right)
\]
where $\bar L_1(m)=L_1(\lambda /(m+1))$.
The expressions of the NNLO
corrections  to the energy levels and
wave functions at the origin 
are rather long \cite{PPres,mel:mass,PY}. 
As an example I give the correction 
$\Delta^{(1)}_2\psi^2_{0m}$ due to the second iteration of 
the $O(\al_s)$ correction to the static potential \cite{direg}
\[
\Delta^{(1)}_2\psi^2_{0m}=\left({\alpha_s\over 4\pi}\right)^2
\left(3(C_0^1+(\bar L(m)+
\Psi_1(m+2))C_1^1)^2\right.
\]
\[
+2C_1^1\left(\sum_{n=0}^{m-1}{(n+1)(m+1)\over (n-m)^3}\left(
C_0^1+\left((\bar L(m)+
\Psi_1(n+2)+{1\over 2}{n+1\over (n-m)(m+1)}\right)C_1^1\right)
\right.
\]
\[
\left.
-\sum_{n=m+1}^{\infty}{(m+1)^2\over (n-m)^3}\left(
C_0^1+\left((\bar L(m)+
\Psi_1(n+2)-{1\over 2}{1\over n-m}\right)C_1^1\right)
\right)
\]
\[
+\left.
2C_1^1\left(C_0^1+\left(\bar L(m)+\Psi_1(m+2)\right)
C_1^1\right)
\left(-{5\over 2}+\sum_{n=0}^{m-1}{n+1\over (n-m)^2}U(m,n)
\right.
\right.
\]
\[
\left.
-\sum_{n=m+1}^{\infty}{m+1\over (n-m)^2}U(m,n)
\right)
+2(C_1^1)^2\left({1\over 2}-\sum_{n=0}^{m-1}{n+1\over (n-m)^2}
+\sum_{n=m+1}^{\infty}{m+1\over (n-m)^2}\right.
\]
\[
\left.
+{1\over 2}\sum_{n=0}^{m-1}{n+1\over (n-m)^3}U(m,n)
+{1\over 2}\sum_{n=m+1}^{\infty}{(m+1)^2\over (n-m)^3(n+1)}U(m,n)
\right.
\]
\[
+\sum_{n=1}^{m-1}\sum_{l=0}^{n-1}\left(
{(l+1)(n+1)\over (n-m)^2(l-m)^2}-{(l+1)(m+1)\over (n-m)^2(l-m)(n-l)}-
{(l+1)(m+1)\over (l-m)^2(n-m)(n-l)}\right)
\]
\[
+\sum_{n=m+1}^{\infty}\sum_{l=0}^{m-1}\left(
-{(l+1)(m+1)\over (n-m)^2(l-m)^2}+{(l+1)(m+1)^2\over (n-m)^2(l-m)(n-l)(n+1)}
\right.
\]
\[
\left.
-{(l+1)(m+1)\over (l-m)^2(n-m)(n-l)}\right)
+\sum_{n=2}^{\infty}\sum_{l=m+1}^{n-1}\left(
{(m+1)^2\over (n-m)^2(l-m)^2}\right.
\]
\[
\left.\left.\left.
+{(l+1)(m+1)^2\over (n-m)^2(l-m)(n-l)(n+1)}+
{(m+1)^2\over (l-m)^2(n-m)(n-l)}\right)\right)\right)
\]
where $\bar L(m)=L(\lambda /(m+1))$ and
\[
U(m,n)=3+{n+1\over m+n+2}
-2{(n+1)^2\over (n-m)(n+m+2)}\, .
\]

\section{Physical results.}
There are basically two application areas
of the near-threshold calculations within NRQCD:
the analysis of $\Upsilon$-resonances 
in the $b\bar b$ system and the $t\bar t$ production cross section near 
the threshold.  

\subsection{$\Upsilon$-resonances.}
For the $b\bar b$ system moments of the spectral density
of different kinds can be studied (cf. eq.~(\ref{mombar0})).
For instance, 
\[
\int_{4m_b^2}^\infty \frac{\rho(s)ds}{s^n}
\]
are meaningful in the near-threshold Coulomb
PT calculations \cite{vol:bres}.
In the NNLO we obtain
the numerical value for the $b$-quark pole mass \cite{PPres}
\be
m_b=4.80\pm 0.06 ~{\rm GeV} .
\label{mbfin}
\ee 
With this result we obtain
the value of the matrix element $|V_{cb}|$
from the analysis of the $B$-meson semileptonic width \cite{PPres} 
\be
\label{final}
|V_{cb}|=0.0423\left({{\rm BR}(B\rightarrow X_cl\nu_l)
\over 0.105}\right)^{1\over 2}
\left({1.55 {\rm ps}\over \tau_B}\right)^{1\over 2}
\left(1 -0.01{\al_s(M_Z)-0.118 \over 0.006}\right)
\left(1\pm\Delta_{npt}\right)
\ee
where $\Delta_{npt}\sim 0.02$
is the uncertainty of nonperturbative contributions.
We use the analysis in the pole mass scheme.
Other approaches based on redefinition of the mass near threshold have
been suggested. The problem of infrared instability of the pole mass 
and its relation to $\MSsch$-mass has been discussed 
\cite{mel:mass,mpoleIRhoa:massrecrenYakM}.

\subsection{$t\bar t$ cross section near threshold.} 

The $t\bar t$-pair near the production 
threshold is just a system that satisfies the requirement of being
nonrelativistic.
Therefore the description of $t\bar t$-system near the production
threshold $\sqrt{s}\approx 2m_t$ ($\sqrt{s}$ is a total energy
of the pair,
$m_t$ is the top quark mass) is quite precise within NRQCD.
The top quark is very heavy $m_t=175~{\rm GeV}$ \cite{PDG}
and there is an energy region of about $8-10$ GeV near the 
threshold where the nonrelativistic approximation
for the kinematics is very precise.
Relativistic effects are small and can be taken into account
perturbatively. The strong coupling constant 
at the high energy scale is small 
$\al_s(m_t)\approx 0.1$ that makes 
the mapping of QCD onto the low energy 
effective theory (NRQCD), which is perturbative in $\al_s(m_t)$,
numerically precise.
The top quark decay width is large, $\G_t=1.43$ GeV;
the infrared (small momenta) region is suppressed and PT calculations 
for the cross section near the threshold are reliable 
even point-wise in energy \cite{FK}.
One can study the $t\bar t$ system near 
the threshold in the processes $e^+e^-\to t\bar t$ \cite{ttee}
and $\gamma\gamma \to t\bar t$ \cite{ttgg}:
\begin{itemize}
\item
$e^+e^-\to t\bar t$: the production vertex is local,
the basic observable is a production cross section
which is saturated by S-wave (for the vector current),
NNLO analysis is available.
\item
$\gamma\gamma \to t\bar t$: the production vertex is 
nonlocal (T-product of two electromagnetic currents),
both S- and P-waves can be studied for different helicity 
photons, the number of observables is larger 
(cross sections $\sigma_S$,
$\sigma_P$, S-P interference).
Because of nonlocality of the production vertex
the high energy coefficient (necessary for mapping the QCD quantities
to NRQCD ones) is more difficult to obtain. 
Some calculations were done in NLO \cite{KMgg,Piv}
and the full analysis of the cross section
is available in NLO of NRQCD only \cite{PP2}. 
The low energy part of the process can be studied in NNLO
without a strict normalization to full QCD (see \cite{INOK} for
relativistic corrections).
\end{itemize}
The top quark width $\G_t$ plays a crucial role
in the calculation of the $t\bar t$ production
cross section near the threshold.
It is taken into account 
by the shift $E\to E+i\G_t$ \cite{FK}.
One has 
\be
\label{dispr}
\sigma(E)\sim {\rm Im}~\Pi (E+i\G_t)
= {\rm Im}~\int \frac{\rho(E')dE'}{E'-E-i\G_t}
= \G_t\int \frac{\rho(E')dE'}{(E'-E)^2+\G_t^2}\, .
\ee
Because the point $E+i\G_t$ lies sufficiently far from 
the positive semiaxis in the complex energy plane
the cross section eq.~(\ref{dispr}) is calculable point-wise in 
energy.
The hadronic cross section $\sigma(E)$ was obtained by many authors
(as a review see, \cite{revHoa}).
\begin{figure}[!ht]
       \epsfig{file=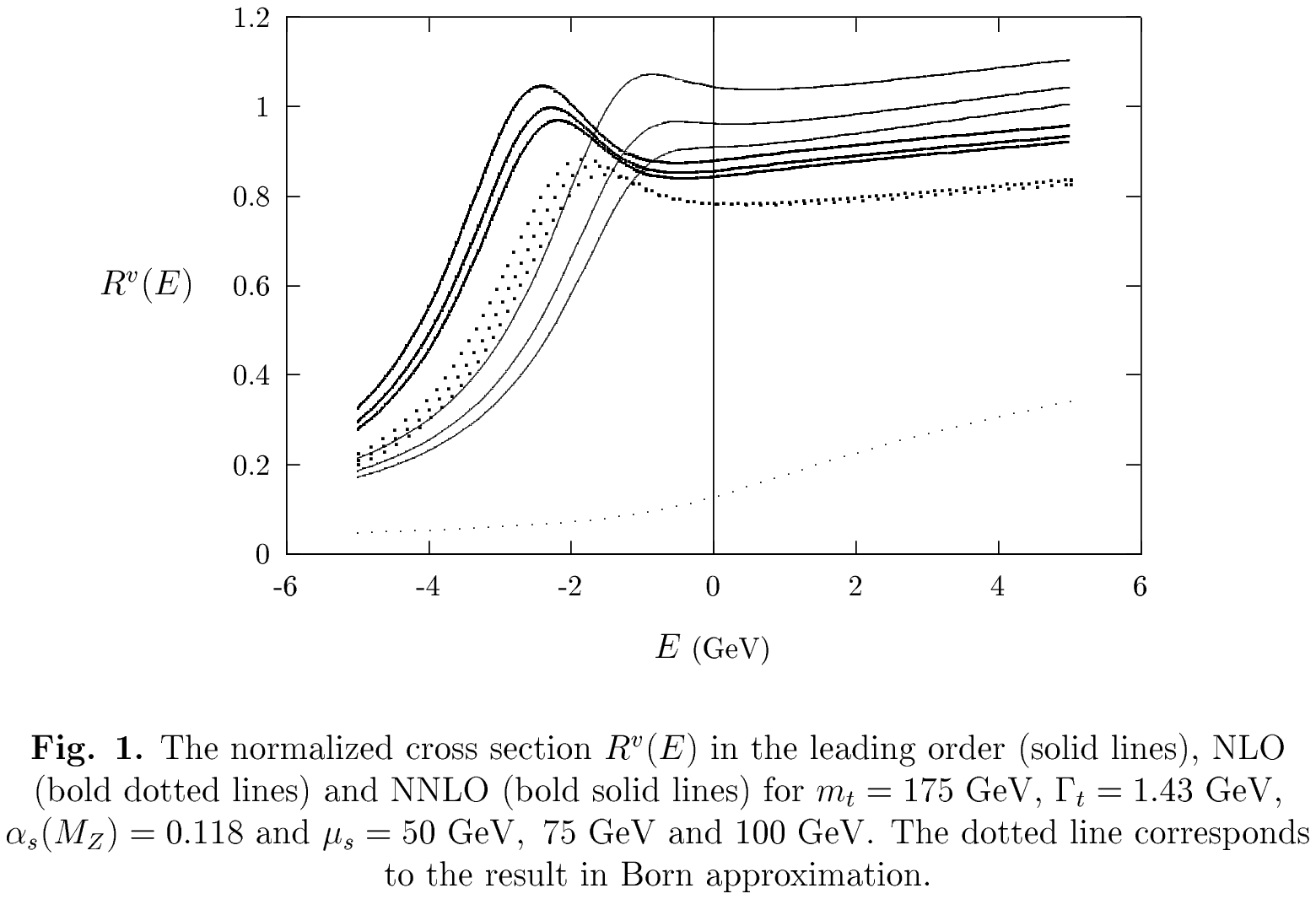,%
angle=0,%
width=1.\textwidth,%
height=0.5\textheight%
}
\end{figure}
The normalized cross sections $R^v(E)$
are plotted in Fig.1 \cite{direg}.
The curves have characteristic points
which are usually considered as basic observables.
They are:
$E_p$ -- the position of the peak in the cross section
and $H_p$ -- the height of the peak.
The convergence for $E_p$ and $H_p$ in consecutive orders 
of PT near the Coulomb solution is not fast in 
the $\MSsch$-scheme \cite{revHoa,PY,Y1}.
For typical values $m_t=175$~GeV, $\G_t=1.43$~GeV, 
$\al_s(M_Z) = 0.118$ one finds \cite{direg}
\bea
E_p&=&E_0(1+0.58+0.38+\ldots)\nn \\
H_p&=&H_0(1-0.15+0.12+\ldots)\, .
\eea 
The way of dealing with the PT expansion of the static potential 
is an important issue in getting stable results within NRQCD
\cite{av}. The convergence in the $\MSsch$-scheme 
is not fast which reflects the
physical situation that the observables 
$E_p$ and $H_p$ are sensitive to different scales. 
The finite-order perturbation theory expansion of the static potential 
given in eq.~(\ref{stapot}) cannot handle several distinct 
scales with the same accuracy. 

\section{Conclusion.}
Some calculations of PT corrections
to systems with heavy particles are presented.
The use of effective theories for computing high order corrections
is crucial in obtaining precise results near the threshold. 
The calculation are rather tedious but still can be completed
analytically in some cases.

\vspace{5mm}
\noindent
{\large \bf Acknowledgments}\\[2mm]
This work is partially supported
by Volkswagen Foundation under contract
No.~I/73611 and Russian Fund for Basic Research under contract
99-01-00091.


\begin{thebibliography}{99}
\bibitem{exp}E. Accomando {\it et al.}, Phys. Rep. {\bf 299} (1998) 1.
\bibitem{weinberggasserleutw}S. Weinberg, Physica {\bf A96} (1979) 327;
J. Gasser, H. Leutwyler, 
Nucl. Phys. {\bf B250} (1985) 465.
\bibitem{shifmanvoloshinpolitzWEinGeorNeub}
M.A. Shifman, M.B. Voloshin,  Sov. J. Nucl. Phys. {\bf 45} (1987) 292;
H.D. Politzer, M.B. Wise, Phys. Lett. {\bf B206} (1988) 681,
Phys. Lett. {\bf B208} (1988) 504;
E. Eichten and B. Hill, Phys. Lett. {\bf B234} (1990) 511;
H. Georgi, Nucl. Phys. {\bf B363} (1991) 301;
M. Neubert, Phys. Rep. {\bf 245} (1994) 259.
\bibitem{isgurwise}N. Isgur, M.B. Wise,
Phys. Lett. {\bf B232} (1989) 113,  Phys. Lett. {\bf B237} (1990) 527.
\bibitem{CasLep} 
W.E. Caswell and G.E. Lepage, Phys. Lett. {\bf B167} (1986) 437;
G.E. Lepage {\it et al.},  Phys.Rev. {\bf D46} (1992) 4052;
G.T. Bodwin, E. Braaten and G.P. Lepage, Phys.Rev. {\bf D51} (1995) 1125.           
\bibitem{revHoa}A.H. Hoang {\it et al.}, Eur. Phys. J.direct  
{\bf C3} (2000) 1.
\bibitem{pivbar}A.A. Ovchinnikov, A.A. Pivovarov, L.R. Surguladze,
Sov. J. Nucl. Phys. {\bf 48} (1988) 358,
Int. J. Mod. Phys. {\bf A6} (1991) 2025.
\bibitem{grobar}S. Groote, J.G.~K\"orner and A.A.~Pivovarov,
Phys. Rev. {\bf D61} (2000) 071501.
\bibitem{polemass}R. Tarrach, Nucl. Phys. {\bf B183} (1981) 384.
\bibitem{sun0}
F.A. Berends, A.I. Davydychev, N.I. Ussyukina, Phys. Lett.
 {\bf B426} (1998) 95;
S.~Groote, J.G.~K\"orner and A.A.~Pivovarov,
Phys. Rev. {\bf D60} (1999) 061701,
Eur.\ Phys.\ J.\  {\bf C11} (1999) 279,
Nucl.\ Phys. {\bf B542} (1999) 515,
Phys.\ Lett. {\bf B443} (1998) 269;
S.~Groote and A.A.~Pivovarov, Nucl. Phys. {\bf B580} (2000) 459; 
A.I.~Davydychev and V.A.~Smirnov, 
Nucl.~Phys.\ {\bf B554} (1999) 391;
N.E. Ligterink, Phys. Rev. {\bf D61} (2000) 105010.
\bibitem{sunappl}J.O. Andersen, E. Braaten, 
M. Strickland, Phys. Rev. {\bf D62} (2000) 045004; 
S.~Narison and A.A.~Pivovarov,
Phys. Lett. {\bf B327} (1994) 341;
T. Sakai, K. Shimizu and K. Yazaki, nucl-th/9912063; 
S.A. Larin {\it et al.}, Sov. J. Nucl. Phys. {\bf 44} (1986) 690;
J.M. Chung, B.K. Chung, Phys. Rev. {\bf D60} (1999) 105001;
K. Chetyrkin, S. Narison, Phys. Lett. {\bf B485} (2000) 145; 
H.Y. Jin, J.G. K\"orner, hep-ph/0003202.
\bibitem{pivan}A.A. Pivovarov, L.R. Surguladze,
Yad. Fiz. {\bf 48} (1988) 1856;
Nucl. Phys. {\bf B360} (1991) 97.
\bibitem{groote}S. Groote, J.G.~K\"orner and O.I.~Yakovlev,
Phys. Rev. {\bf D55} (1997) 3016.
\bibitem{barmatch}A.G.~Grozin and O.I.~Yakovlev,
Phys. Lett. {\bf 285B} (1992) 254.
\bibitem{chetkwiaschw3}K.G. Chetyrkin, J.H. K\"uhn, A. Kwiatkowski,
Phys. Rept. {\bf 277} (1996) 189;
K.G. Chetyrkin, J.H. K\"uhn, M. Steinhauser,
Phys. Lett. {\bf B371} (1996) 93, Nucl. Phys. {\bf B482} (1996) 213.
\bibitem{BetheSal}E.E. Salpeter, H.A. Bethe, 
Phys. Rev. {\bf 84} (1951) 1232.
\bibitem{VL}M.B. Voloshin, Nucl. Phys. {\bf B154} (1979) 365;
Yad. Fiz. {\bf 29} (1979) 1368, Sov. J. Nucl. Phys. {\bf 36} (1982) 143;
H. Leutwyler, Phys. Lett. {\bf B98} (1981) 447.
\bibitem{ttee}W. Kwong, Phys. Rev. {\bf D43} (1991) 1488;
M.J. Strassler and M.E. Peskin,  Phys. Rev. {\bf D43} (1991) 1500;
M. Jezabek, J.H. K{\"u}hn and T. Teubner, Z. Phys. {\bf C56} (1992) 653;
Y. Sumino {\it et al.}, Phys. Rev. {\bf D47} (1993) 56.
\bibitem{KarHarBr}R. Karplus and A. Klein,  
Phys. Rev. {\bf 87} (1952) 848;
G. K{\"a}llen and A. Sarby,
K. Dan. Vidensk. Selsk. Mat.-Fis. Medd. {\bf 29} (1955) , N17, 1;
R. Barbieri {\it et al.}, Phys. Lett. {\bf B57} (1975) 535;
I. Harris and L.M. Brown, Phys. Rev. {\bf 105} (1957) 1656;
R. Barbieri {\it et al.}, Nucl. Phys. {\bf B154} (1979) 535.
\bibitem{hoa:match}A.H. Hoang, Phys.Rev. {\bf D56} (1997) 7276.
\bibitem{highcoef}A. Czarnecky and K. Melnikov,
Phys. Rev. Lett. {\bf 80} (1998) 2531;
M. Beneke, A. Signer and V.A. Smirnov, 
Phys. Rev. Lett. {\bf 80} (1998) 2535.
\bibitem{pivKK}A.A. Pivovarov, 
Phys. Lett. {\bf B236} (1990) 214, Phys. Lett. {\bf B263} (1991) 282.
\bibitem{Fish}W. Fisher, Nucl. Phys. {\bf B129} (1977) 157;
A. Billoire, Phys. Lett. {\bf B92} (1980) 343.
\bibitem{PeterSchr}M. Peter, Phys. Rev. Lett. {\bf 78} (1997) 602;
Nucl. Phys {\bf B501} (1997) 471;
Y. Schr{\"o}der, Phys. Lett. {\bf B447} (1999) 321.
\bibitem{AshBerLandau}A.I. Achieser and V.B. Berestezki,
Quantum electrodynamics, Moscow 1959;
L.D. Landau and E.M. Lifshitz, 
Relativistic Quantum Theory, Part 1 (Pergamon, Oxford, 1974) .
\bibitem{KPP}J.H. K{\"u}hn, A.A. Penin and A.A. Pivovarov,
Nucl. Phys. {\bf B534} (1998) 356.
\bibitem{PPres}A.A. Penin and A.A. Pivovarov, 
Phys. Lett. {\bf B435} (1998) 413; Phys. Lett. {\bf B443} (1998) 264; 
Nucl. Phys. {\bf B549} (1999) 217.
\bibitem{BSS1}M. Beneke, A. Singer and V.A. Smirnov,
Phys. Lett. {\bf B454} (1999) 137.      
\bibitem{Hoang}A.H. Hoang and T. Teubner,
Phys.Rev. {\bf D58} (1998) 114023.
\bibitem{Mel}K. Melnikov and A. Yelkhovsky, 
Nucl. Phys. {\bf B528} (1998) 59.
\bibitem{Nag}T. Nagano, A. Ota and  Y. Sumino, 
Phys. Rev. {\bf D60} (1999) 114014.
\bibitem{direg}A.A. Penin and A.A. Pivovarov, 
MZ-TH-98-61, Dec 1998. 41pp. [hep-ph/9904278].
\bibitem{Yak}O. Yakovlev, Phys. Lett. {\bf B457} (1999) 170.
\bibitem{HoaTeu}A.H. Hoang and T. Teubner,   
Phys. Rev. {\bf D60} (1999) 114027.
\bibitem{nonrelPiv}A.A. Pivovarov, Phys. Lett. {\bf B475}(2000) 135.
\bibitem{PP2}
A.A. Penin and A.A. Pivovarov, Nucl. Phys. {\bf B550} (1999) 375.
\bibitem{mel:mass} 
K.Melnikov and A. Yelkhovsky, Phys. Rev. {\bf D59} (1999) 114009.
\bibitem{PY} A. Pineda and F.J. Yndurain, Phys.Rev. {\bf D58} (1998) 094022.
\bibitem{vol:bres}M. Voloshin, Int. J. Mod. Phys. {\bf A10} (1995) 2865.
\bibitem{mpoleIRhoa:massrecrenYakM}
M. Beneke and V.M. Braun, Nucl. Phys. {\bf B426} (1994) 301;
I.I. Bigi {\it et al.}, Phys.Rev. {\bf D50} (1994) 2234;
A.H. Hoang, Phys. Rev. {\bf D59} (1999) 014039;
K.G. Chetyrkin, M. Steinhauser, 
Phys. Rev. Lett. {\bf 83} (1999) 4001;
S. Groote and O. Yakovlev, hep-ph/0008156.
\bibitem{PDG}Particle Data Group, Eur.~Phys.~J.\ {\bf C3} (1998) 1.
\bibitem{FK}V.S. Fadin and V.A. Khoze, Pis'ma Zh. Eksp. Teor. Fiz.
{\bf 46} (1987) 417; Yad. Fiz. {\bf 48} (1988) 487.
\bibitem{ttgg}V.S. Fadin and V.A. Khoze, Yad.Fiz. {\bf 93} (1991) 1118;
I.I. Bigi, V.S. Fadin  and V.A. Khoze, Nucl. Phys. {\bf B377} (1992) 461;
I.I. Bigi, F. Gabbiani and V.A. Khoze, Nucl. Phys. {\bf B406} (1993) 3;
J.H. K\"uhn, E. Mirkes and J. Steegborn, Z. Phys. {\bf C57} (1993) 615. 
\bibitem{KMgg}B. Kamal, Z. Merebashvili and A.P. Contogouris,  
Phys.Rev. {\bf D51} (1995) 4808.
\bibitem{Piv}A.A. Pivovarov, Phys.Rev. {\bf D47} (1993) 5183.
\bibitem{INOK}K.A. Ispiryan {\it et al.}, Yad.Fiz. {\bf 11} (1970) 1278.
\bibitem{Y1}F.J. Yndurain, hep-ph/0007333, hep-ph/0008007.
\bibitem{av}M. Jezabek, J.H.K{\"u}hn, M.Peter, 
Y.Sumino and T. Teubner, Phys. Rev. {\bf D58} (1998) 014006. 



\end{thebibliography}
\end{document}